# Acoustomicrofluidic separation of tardigrades from raw cultures for sample preparation

Muhammad Afzal[1,#], Jinsoo Park[1,#], Ghulam Destgeer[1,#], Husnain Ahmed[1], Syed Atif Iqrar[1], Sanghee Kim[2], Sunghyun Kang[3], Anas Alazzam[4], Tae-Sung Yoon[3,*] and Hyung Jin Sung[1,*]

[1]Department of Mechanical Engineering, KAIST, Daejeon 34141, Korea

[2]Division of Polar Life Sciences, KOPRI, Incheon 21990, Korea

[3]Department of Proteome Structural Biology, KRIBB, Daejeon 34141, Korea

[4]Department of Mechanical Engineering, Khalifa University, Abu Dhabi 127788, UAE

[#]These authors contributed equally to this work.

*Corresponding authors. E-mail: hjsung@kaist.ac.kr and yoonts@kribb.re.kr

## Abstract

Tardigrades are microscopic animals widely known for their survival capabilities under extreme conditions. They are the focus of current research in the fields of taxonomy, biogeography, genomics, proteomics, development, space biology, evolution, and ecology. Tardigrades, such as *Hypsibius exemplaris*, are being advocated as a next-generation model organism for genomic and developmental studies. The raw culture of *H. exemplaris* usually contains tardigrades themselves, their eggs, and algal food and feces. Experimentation with tardigrades often requires the demanding and laborious separation of tardigrades from raw samples to prepare pure and contamination-free tardigrade samples. In this paper, we propose a two-step acousto-microfluidic separation method to isolate tardigrades from raw samples. In the first step, a passive microfluidic filter composed of an array of traps is used to remove large algal clusters in the raw sample. In the second step, a surface acoustic wave-based active microfluidic separation device is used to continuously deflect tardigrades from their original streamlines inside the microchannel and thus selectively isolate them from algae and eggs. The experimental results demonstrated the efficient tardigrade separation with a recovery rate of 96% and an algae impurity of 4% on average in a continuous, contactless, automated, rapid, biocompatible manner.

# 1. INTRODUCTION

Tardigrades, microscopic animals found on all continents, are widely known for their cryptobiotic abilities to survive under extreme physical conditions (Guidetti, Altiero, & Rebecchi, 2011; Møbjerg *et al.*, 2011) such as temperature (Hengherr *et al.*, 2009a,b), pressure (Seki & Toyoshima, 1998; Ono *et al.*, 2008, 2016), space vacuum (Jönsson *et al.*, 2008), UV (Altiero *et al.*, 2011; Horikawa *et al.*, 2013; Giovannini *et al.*, 2018), and Gamma ray exposure (Beltrán-Pardo *et al.*, 2015). Much effort has been devoted to characterizing the taxonomy (e.g. Kosztyła *et al.*, 2016; Jørgensen, Kristensen, & Møbjerg, 2018), biogeography (e.g. Pilato & Binda, 2001), evolution and ecology (e.g. Nelson, 2002), genomics (e.g. Yoshida *et al.*, 2017), proteomics (e.g. Schokraie *et al.*, 2012), development (e.g. Smith *et al.*, 2016; Gross, Minich, & Mayer, 2017), and space biology (e.g. Rebecchi *et al.*, 2009; Weronika & Łukasz, 2017) of tardigrades in an effort to better understand this group of microscopic organisms. Recently, the recovery and reproduction of a tardigrade that had been frozen for over 30 years attracted a great deal of attention (Tsujimoto, Imura, & Kanda, 2016). Tardigrades have been found to produce specific proteins (e.g., CAHS, SAHS, MAHS, RvLEAM) under external stresses to enable survival (Yamaguchi *et al.*, 2012; Tanaka *et al.*, 2015). A unique DNA-associated protein (Dsup) from the tardigrade has been expressed in the human cultured cells to improve human radiation tolerance and reduce X-ray-induced damage of the cellular DNA by 40% (Hashimoto *et al.*, 2016).

With the aforementioned characteristics, tardigrades have been suggested as potential model organisms for space research (Jönsson, 2007) as well as for developmental and genomic studies to be investigated along with two commonly used model organisms: *Caenorhabditis elegans* (Nematoda) and *Drosophila melanogaster* (Arthropoda) (Tenlen, McCaskill, & Goldstein, 2013; Goldstein & King, 2016; Martin *et al.*, 2017). The *Hypsibius exemplaris* (Gasiorek *et al.*, 2018) tardigrade has been specifically investigated as a new model animal for evolutionary developmental research due to its characteristics shared with the two model animals mentioned above, short generation time and ease to culture continuously for decades (Gabriel *et al.*, 2007). The genome of *H. exemplaris* has been sequenced to investigate the evolution of molecular and developmental mechanisms, and it was erroneously suggested that a horizontal gene transfer from other animals has undesirably impacted the genome composition of the tardigrade *H. exemplaris* (Boothby *et al.*, 2015). However, some follow-up studies by other researchers refuted this earlier claim and contend that the false finding was attributed to bacterial contamination derived from the uncontrolled sample preparation (Bemm *et al.*, 2016; Koutsovoulos *et al.*, 2016; Arakawa, 2016). To prevent false-positive results in the experimental data, it is imperative to develop effective separation methods for selective isolation of tardigrades from raw cultures.

Increased research interest in tardigrades has demanded advancements in supportive technologies to facilitate experimentation and the development of standard protocols. Several studies on the laboratory-scale culture methods of tardigrades have been reported (Altiero & Rebecchi, 2001; Suzuki, 2003; Horikawa *et al.*, 2008; Tsujimoto, Suzuki, & Imura, 2015; Altiero *et al.*, 2015). The growth cultures of tardigrades contain different age groups of tardigrades (adults, newborns, and juveniles of variable sizes), eggs, exuviae, and algal food. It is important to prepare homogenous samples of tardigrades having a uniform age or size distribution for further experimentation (Gabriel *et al.*, 2007). Separating tardigrades from food, eggs, exuviae, or unwanted debris from a raw sample is laborious and time-consuming work, in which individual tardigrades must be manually picked using a wire Irwin loop (Sands *et al.*, 2008), a micropipette (Degma, Katina, & Sabatovičová, 2011), or a needle (Gross *et al.*, 2017) under a microscope. These labor-intensive methods require a skilled person to dexterously locate and capture a single tardigrade within a sample under a microscope, provide a limited separation throughput, and are susceptible to contamination. On the other hand, tardigrades can also be isolated in batches from their algal food using the Baermann filtration technique, in which raw cultures are repeatedly passed through a filter paper (Koutsovoulos *et al.*, 2016). Despite the simple operation and low cost of the filtration method, it has been reported to have inherent limitations such as long processing time and its dependency on sample volume (Van Bezooijen, 2006). The manual and filtration methods described above do not offer continuous, automated, non-contact, on-demand control over the separated constituents.

In recent years, microfluidic approaches to separate microscale objects have been proven to be effective and shown great potential for many medical, biological, chemical and industrial applications (Sajeesh & Sen, 2014; Wyatt Shields IV, Reyes, & López, 2015). Microscale flows within a microchannel allows precise control over the flow and suspended objects and continuous sample processing. A variety of microfluidic separation techniques (Bhagat *et al.*, 2010), including passive methods that utilize hydrodynamics forces arising from the microchannel geometry and active methods that rely on external force fields, have been reported to offer rapid, continuous, automated, biocompatible sample processing. Inertial microfluidics (Hur, Tse, & Di Carlo, 2010), hydrodynamic filtration (Choi *et al.*, 2008), magnetophoresis (Alnaimat *et al.*, 2018), optofluidics (Jung *et al.*, 2014), dielectrophoresis (Mathew *et al.*, 2016), and acoustofluidics (Ding *et al.*, 2013; Destgeer & Sung, 2015) are common techniques for isolating micro-objects based on size, magnetic, optical, electrical, or mechanical properties. In particular, acoustofluidic platforms are widely used for numerous applications, including cell sorting (Ma *et al.*, 2017), microparticle separation (Ahmed *et al.*, 2017a, 2018), microparticle patterning (Destgeer *et al.*, 2016, 2019; Collins *et al.*, 2018), microscale flow mixing (Nam, Jang, & Lim, 2018; Ahmed *et al.*, 2019), chemical and thermal gradient generation (Destgeer *et al.*, 2014b; Ha *et al.*, 2015), protein isolation (Ahmad *et al.*, 2017), droplet handling (Park *et al.*, 2017a, 2018b), in-droplet microparticle manipulation (Park *et al.*, 2017b, 2018a), nebulization (Winkler *et al.*, 2017), and microorganism manipulation (Ding *et al.*, 2012). Microfluidic platforms can be utilized for selective isolation of tardigrades from complex samples to address the limitations of the conventional tardigrade separation techniques based on pipettes, sieves, filter papers, or Irwin loops.

In this paper, we have combined passive and active microfluidic separation techniques to isolate *H. exemplaris* tardigrades from raw samples. In the first step, a laboratory culture of tardigrades was passed through a passive microfluidic filter to capture large algal clusters. The passive filter was composed of an array of micro-pillars that acted as traps and were built inside a polydimethylsiloxane (PDMS) microchannel to allow passage of micro-objects (i.e., tardigrades, eggs, algae) smaller than the trap pore size while trapping larger objects (i.e., algal clusters). The sample collected from the outlet of the passive filter contained tardigrades, exuviae, eggs, and algal food, and it was then passed through an active separation device for the second step. The active microfluidic platform was composed of an interdigital transducer (IDT) mounted on top of a piezoelectric substrate and a PDMS microfluidic chip. Surface acoustic waves (SAWs) produced from the IDT were transformed into compressional waves (CWs) inside the microchannel and exerted an acoustic radiation force (ARF) on the tardigrades in the direction of wave propagation. The deflected tardigrades were thereby isolated from any remaining algae, exuviae, and eggs with the achievement of 96% of recovery rate and 4% of algae impurity. The experimental demonstration has proven that the proposed method is a promising technique for tardigrade sample preparation in a continuous, automated, rapid, biocompatible, non-contact, on-demand manner.

## 2. MATERIAL AND METHODS

### 2.1 TARDIGRADE CULTURE

Cultures containing the tardigrade *Hypsibius exemplaris* and *Chlorococcum* sp. algae were purchased and grown according to the protocol provided by the supplier (Sciento). The *H. exemplaris* tardigrades were grown in Chalkley's medium plus soil extract at 16 °C, with light and dark periods of 14:10 hours (Arakawa, Yoshida, & Tomita, 2016). The *Chlorococcum* algae, used as tardigrade food, were prepared in Bold's Basal medium plus soil extract at 25 °C under 16:8 hours of light and dark periods. Every 4 to 6 weeks, subcultures of the tardigrades were prepared, as they had consumed their algal food. On average, a new generation of tardigrades was produced every two weeks. Tardigrades laid eggs inside their exuviae. The exuviae of a healthy tardigrade may contain up to 36 eggs that hatch within 4–5 days. The juveniles matured into adult tardigrades within 2 weeks, at which point they were ready to lay eggs (Gabriel *et al.*, 2007).

## 2.2 TARDIGRADE SEPARATION PROCESS

Figure 1 shows the workflow of the two-step tardigrade separation process composed of the passive filtration of algal clusters and the active separation of tardigrades from eggs and algae. Unicellular *Chlorococcum* algae tend to flocculate to form large clusters in the tardigrade growth cultures. These aggregated clusters even exceed tardigrades in size; thus, they often cause microchannel clogging and produce pulsating fluctuations that significantly disturb the laminar flow inside the microfluidic channel. For these reasons, the algal clusters present in the growth cultures should be removed for the effective separation of tardigrades prior to the acoustofluidic tardigrade separation. In the first step, the raw sample was passively filtered in the passive microfluidic device so that the large clusters of algae were removed by an array of algal cluster traps, each of which consisted of four micropillars. As a result, the amount of algae present in the raw tardigrade sample could be significantly decreased. The micropillars of the traps also broke the flocculated algal clusters into smaller units to facilitate the active microfluidic tardigrade separation. In the second step, the filtered sample was continuously pumped into an active separation device to selectively separate the tardigrades from the remaining algae and eggs. The SAW-induced ARF acting on the tardigrades deflected them from their original laminar streamlines inside a microchannel and thus sorted them into a desired outlet while the algae and eggs were collected at a different outlet without being affected by the acoustic field. The detailed underlying physics of selective separation of tardigrades will be discussed in detail in below.

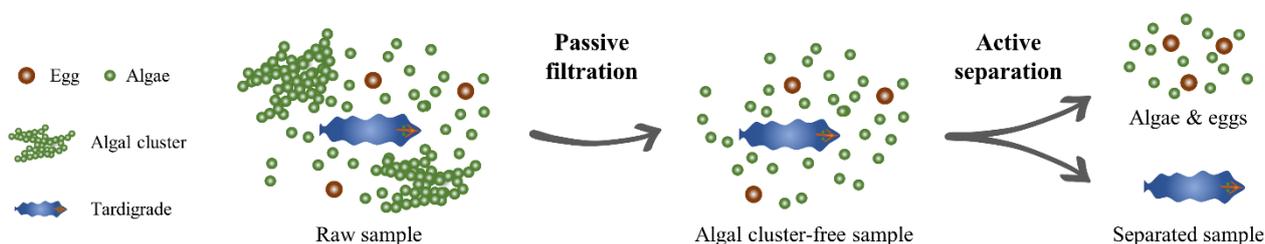

**Figure 1.** A schematic diagram showing the overall workflow of the two-step acousto-microfluidic separation of tardigrades. In the first step, algal clusters in the tardigrade growth culture were filtered by an array of traps composed of micropillars. In the second step, tardigrades were selectively isolated from any remaining algae and eggs for pure sample preparation in an active separation device.

## 2.3 DEVICE DESIGN AND FABRICATION

Passive and active microfluidic devices were separately designed and fabricated in parallel prior to conducting the experiments in series. A passive filtration device was designed to have a total of 218 algal cluster traps composed of four micropillars of 200 μm in diameter. The pillars were arranged in a polar array having a 500 μm radius, 40° angle, and 142 μm distance between adjoining pillars, which determined the pore size of the traps, as shown in Figure 2. A soft lithography technique was used to fabricate PDMS microfluidic chips on which a microchannel is patterned (Friend & Yeo, 2010; Destgeer *et al.*, 2013). PDMS has been chosen as the microchannel material because it is biocompatible, optically transparent, gas-permeable and chemically stable (Weibel *et al.*, 2006). A negative photoresist (SU-8 2150, MicroChem), coated on a silicon wafer using a spin coater (Midas System) and sequentially baked on hot plates (65 °C and 95 °C, respectively), was selectively exposed to UV through a chrome/glass mask (NEPCO) mounted on the stage of a mask aligner (Midas System). The height of the algal cluster traps in the passive microfluidic filter was defined as the thickness of the photoresist layer. After baking, the wafer was placed in a SU-8 developer solution to remove the uncured photoresist and thus obtain the desired microchannel pattern. The PDMS base and curing agent (Sylgard 184A and 184B, Dow Corning) were thoroughly mixed with a weight ratio of 10:1, and the PDMS mixture was poured onto the patterned SU-8 mold on the silicon wafer placed inside a square petri dish. The petri dish was placed in a vacuum chamber and degassed for bubble removal. After the PDMS solution was cured in an oven at 65 °C for two hours, PDMS microfluidic chips were peeled off the Si substrate and cut into appropriate sizes. The inlet and outlet ports were punched (Harris Uni-Core) through the microchannels for sample injection and collection. The surfaces of the PDMS microchannels and a glass slide were treated with oxygen plasma for 2 minutes (Covance, Femto Science) and gently placed in contact to achieve irreversible bonding.

The active tardigrade separation device consisted of an IDT deposited on a piezoelectric substrate and a PDMS microfluidic chip, on which a desired microchannel was patterned. The IDTs were fabricated by spin-coating a photoresist layer onto a 128° rotated Y-cut, X-propagating $LiNbO_3$ substrate (MTI Korea) and patterning the resist in the reverse shape of comb-like transducers. A bimetallic conductive layer composed of a 30 nm thick layer of chrome that promoted adhesion of a 1000 Å thick gold layer was deposited by the e-beam evaporation process onto the piezoelectric substrate, followed by a lift-off process to remove the excess metals and photoresist. Two IDTs with uniform electrode width and spacing were formed on the substrate. Electrodes having width and spacing of 20 μm and 12.5 μm were designed to obtain IDTs with resonant frequencies at 45 MHz and 72 MHz, respectively. A 200 nm-thick $SiO_2$ layer was deposited to cover the electrodes to protect them from mechanical damage (Destgeer *et al.*, 2015). PDMS microfluidic chips were fabricated using the soft lithography process described above. A thin PDMS membrane was used to seal the PDMS microfluidic chip to investigate the effects of the SAW-induced ARF on tardigrades, as shown in Figure 3 (Park *et al.*, 2018b). The 38 μm-thick PDMS membrane was prepared by spin-coating the base and curing agent mixture on top of the saline-treated silicon wafer. After oxygen plasma treatment, the PDMS microchannel and membrane were permanently bonded. The height and width of the PDMS microfluidic channel were 190 μm and 500 μm, respectively. The PDMS microchannel was placed directly on top of the IDT (72 MHz) to reversibly bond it to the $LiNbO_3$ substrate covered with a $SiO_2$ layer. For tardigrade separation, a PDMS microfluidic chip with 163 μm in height and 500 μm in width was used; this microchannel was not sealed using a PDMS membrane but permanently bonded to a $SiO_2$-covered $LiNbO_3$ substrate such that the IDT (45 MHz) was positioned outside of the microchannel (see Figure 4).

## 2.4 EXPERIMENTAL SETUP

Initially, the raw tardigrade sample was homogenized by repeatedly (8 times) pumped back and forth inside a 15 mL conical centrifuge tube (Falcon). The 2.5 mL tardigrade sample was then injected into the passive microfluidic filtration device using PTFE tubing (0.56 mm ID × 1.07 mm OD, Cole Parmer Company) attached to a gas-tight glass syringe (Hamilton Company) connected with a 27G stainless steel hypodermic needle. After repeated passive filtration, the processed sample was loaded into a separate syringe and injected into the active microfluidic tardigrade separation device using a syringe pump (neMESYS, CETONI GmbH) along with two sheath flows (pure DI water). The SAW-based active separation device was mounted on top of a microscope stage (BX53, Olympus), and a 10-bit high-speed COMS camera (pco.1200 hs, PCO) was used to capture the images. An RF signal generator (N5181A, Agilent Technologies) and an amplifier (LZT-22+, Mini-Circuits) were used to produce high-frequency (45 MHz and 72 MHz) AC signals with varying amplitude to actuate the IDTs. ImageJ software (http://imagej.nih.gov/ij) was used to analyze experimental images, as shown in Figures 3 and 4.

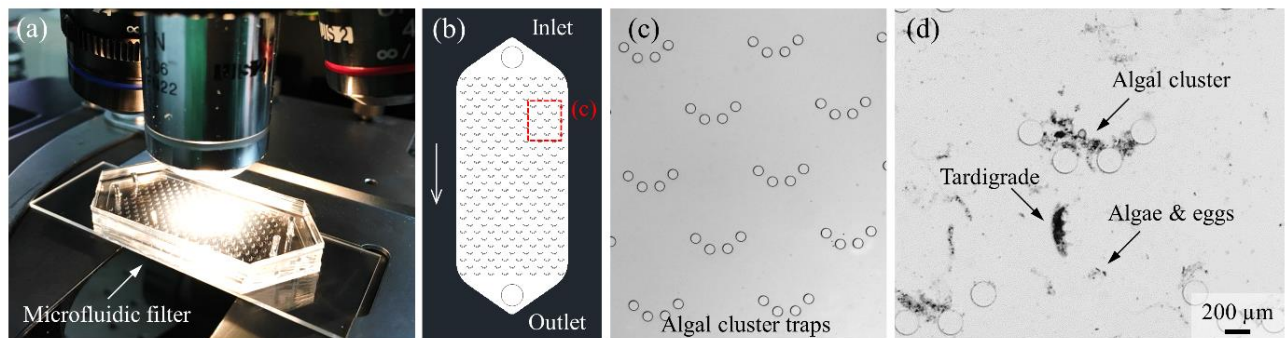

**Figure 2.** A passive filtration process for algal cluster removal from the raw tardigrade sample prior to active tardigrade separation. (a) A PDMS microfluidic filtration device placed on a microscope. (b) The microchannel layout. (c) A magnified view of algal cluster traps composed of four micropillars. (d) A microscopic snapshot in which algal clusters were filtered by the trap whereas a tardigrade passed through it along with smaller algae and eggs.

## 3. RESULTS

### 3.1 PASSIVE FILTRATION

Raw cultures of tardigrades contain numerous algal clusters along with the tardigrades, their eggs, and algae. For effective and efficient acoustofluidic separation of tardigrades, the algal clusters should be removed by passive microfluidic filtration so that the amount of algae present in the sample is significantly decreased. As shown in Figures 2(a–c), the passive microfluidic filter was designed to have an array of 218 traps composed of four micropillars to capture the large algal clusters whereas the tardigrades, their eggs, and algal food readily passed through the gaps between the pillars. The raw tardigrade sample including algal clusters, tardigrades, eggs and algae was injected into the passive microfluidic device at a flow rate of 15–20 mL min⁻¹ (7.5–10 s per round). The distance between the adjacent pillars was 142 μm, so the tardigrades (50–100 μm in length), eggs (40–50 μm in diameter), and algae (10–20 μm in diameter) were allowed to pass through the algal cluster traps, as shown in Figure 2(d). Once a trap was filled with an algal cluster, the increased flow resistance caused the following fluid streams carrying large algal clusters and tardigrades to be diverted away from the occupied trap toward the empty traps downstream of the device. For validation of the algal cluster removal, we measured the optical density (OD) of the tardigrade sample before and after the passive microfluidic filtration by a spectrophotometer (GeneQuant 1300) at 600 nm; the measured OD value of the tardigrade sample was significantly decreased by 76.53% ± 2.88% from 2.198 ± 0.177 to 0.512 ± 0.037 after the algal cluster filtration on average for ten independent repeated experiments.

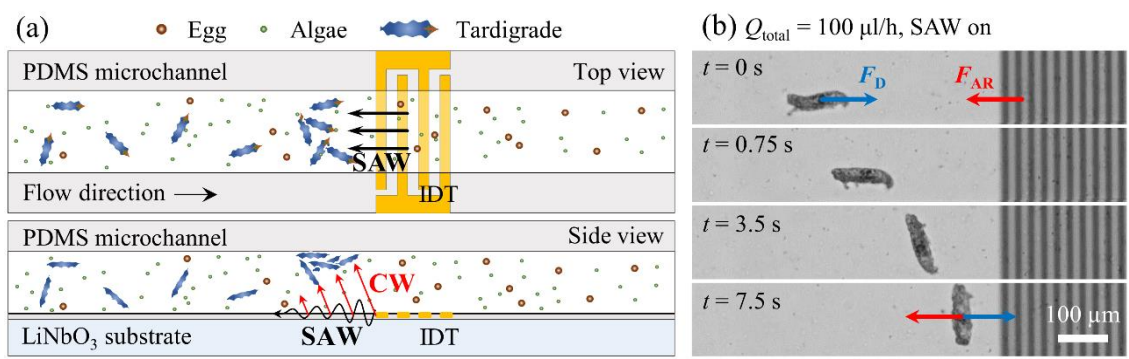

**Figure 3.** (a) A schematic diagram showing the top and side views of the active acoustofluidic platform in which incoming tardigrades were trapped besides the IDT by the SAW-derived ARF ($F_{AR}$). (b) A flow rate of 100 μL h$^{-1}$ exerted a drag force ($F_D$) on the tardigrade, which was balanced by the SAW-based ARF at an equilibrium position.

## 3.2. ACOUSTIC RADIATION FORCE ON TARDIGRADES

For effective acoustofluidic tardigrade isolation, the tardigrades should be selectively affected by SAW-based ARF while the eggs and algae remain unaffected. We investigated the effects of the SAW-induced ARF on the filtered tardigrade sample using a parallel-type acoustofluidic device (Park *et al*., 2017, 2018b), where the wave propagation direction was opposite to the flow direction. Figure 3(a) shows the top and side views of the device, in which the filtered sample was injected through the inlet of the microchannel. The bottom of the microchannel was sealed by a thin PDMS membrane to prevent the tardigrade sample was not in direct contact with the IDT, which may induce undesired side effects by the electric field formed around the electrodes. When AC signals were applied to the IDT at its resonant frequency, SAWs were produced from the electrodes and immediately transformed into CWs. Figure 3(b) shows a series of experimental images where the passively filtered tardigrade sample fluid was continuously passing through the microchannel at a flow rate of 100 µL h⁻¹ when 72 MHz SAWs at 17.2 V$_{pp}$ were produced from the IDT placed underneath the microchannel. The experiments were conducted in the parallel-type device to confirm that the SAW-based ARF was selectively effective to the tardigrades for isolation them from the eggs and algae. As the tardigrade approached the IDT, the magnitude of the ARF acting on the tardigrade was increased until they were located at the equilibrium position of two counter-acting forces: the flow-induced drag force ($F_D$) and ARF ($F_{AR}$), as shown in the figure ($t = 7.5$ s). The vertical component of the ARF continuously pushed the tardigrade toward the ceiling of the microchannel while the horizontal component of the ARF matched the drag force (Ahmed *et al.*, 2017a,b, 2018). In contrast to the tardigrade trapped at the equilibrium position right next to the IDT, the eggs and algae present in the sample passed freely through the microchannel, verifying our hypothesis that the SAW-induced ARF can solely separate the tardigrades from the eggs and algae (see also Supplementary Movie 1). We observed that the tardigrades collected after being exposed to the acoustic field were actively moving (see also Supplementary Movie 2) and succeeded in reproduction.

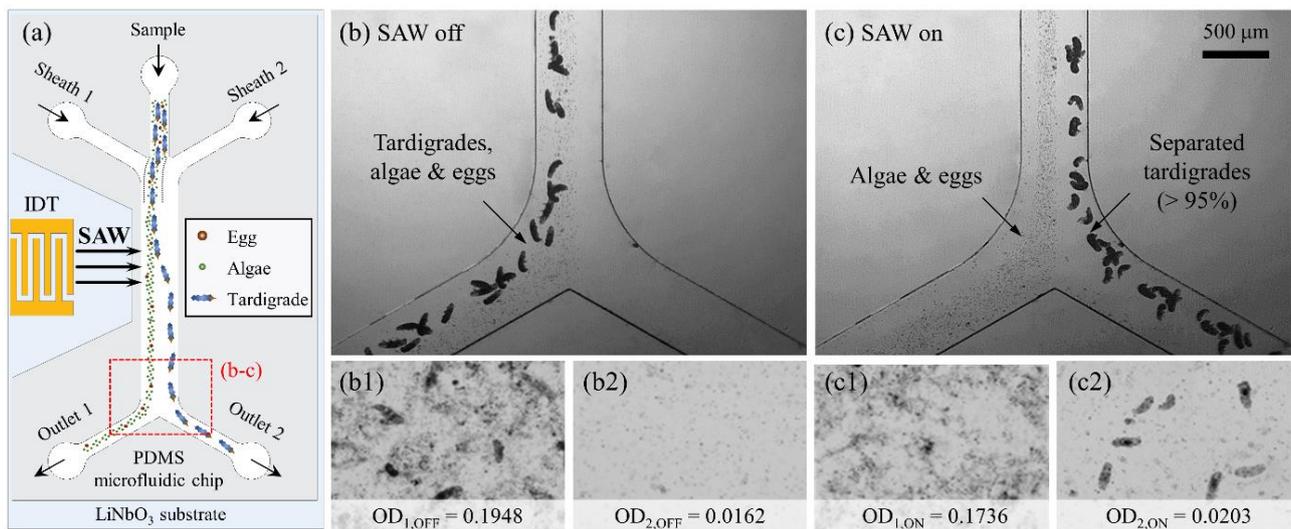

**Figure 4.** (a) A schematic diagram illustrating the continuous separation of tardigrades from algal food and eggs based on SAW-induced ARF. The sample fluid containing tardigrades, their eggs, and algal food was pumped through a central inlet and focused by two sheath flows. The SAW-based ARF deflected the tardigrades away from their original path to separate into a desired outlet port. The eggs and algae were unaffected by the force and thus move along with their original streamlines. (b) When the SAWs were not applied, a focused culture sample fluid passed through the microchannel and was collected at outlet 1 without deflection of the tardigrades. Images (b1) and (b2) of the collected samples at outlets 1 and 2 revealed the presence and absence of tardigrades as well as optical density (OD) values for algae quantification at respective outlets. (c) The ARF produced by 45 MHz SAWs originating from the IDT pushed the tardigrades from their original path to sort them into outlet 2 with a recovery rate of more than 95%. The algae and eggs were mostly unaffected

by the force and thus collected at outlet 1. Images (c1) and (c2) of the outlet samples revealed the absence and presence of tardigrades and OD values for algae impurity quantification at outlets 1 and 2, respectively.

## 3.3. ACOUSTOFLUIDIC SEPARATION OF TARDIGRADES

Figure 4(a) shows a schematic diagram of the cross-type acoustofluidic device, where the wave propagation direction was perpendicular to the flow direction, used to prepare pure samples of tardigrades from raw cultures. The filtered tardigrade sample fluid was injected into the central inlet, along with the sheath flows 1 (left) and 2 (right) of DI water from the two other inlets, to hydrodynamically focus the sample flow at the desired location. The flow rates were 2000 μL h$^{-1}$ (sample flow), 1000 μL h$^{-1}$ (sheath 1), and 3000 μL h$^{-1}$ (sheath 2). The sheath flow 2 from the right-hand inlet was used to pinch the sample flow close to the microchannel sidewall while the sheath flow 1 was introduced to prevent the suspended constituents in the tardigrade sample from being located in the anechoic region, where the effective acoustic field was not formed (Destgeer *et al.*, 2015). The IDT was placed beside the microchannel within an acoustic window to minimize acoustic wave damping so that the acoustic waves could be coupled with the fluid inside the microchannel in an energy-efficient manner (Shi *et al.*, 2009; Destgeer *et al.*, 2013). The SAWs radiating from the IDT propagated along the surface of the piezoelectric substrate and penetrated into the microchannel passed through the narrow PDMS wall with a thickness of 100 μm. Figure 4(b) shows the stacked experimental image in which the hydrodynamically focused sample fluid was collected at the outlet 1 (left) without the acoustic field applied to the microchannel. The microscopic images of the samples collected from the outlets 1 and 2 clearly show the presence (Figure 4(b1)) and absence (Figure 4(b2)) of the tardigrade sample including the tardigrades, eggs and algae. The average OD values of the samples collected at both outlets were measured to be 0.1948 (outlet 1) and 0.0162 (outlet 2), confirming our experimental observation of all the components of the tardigrade sample flowing into the outlet 1. On the other hand, Figure 4(c) shows the stacked microscopic image when 45 MHz SAWs at 17.2 V$_{pp}$ were applied to the microchannel. As clearly seen in the figure, the SAW-induced ARF acting on the tardigrades exclusively deflected them from the original streamlines so that the tardigrades were transferred from the filtered sample to the sheath flow 2. Consequently, the tardigrades could be separated into the outlet 2 whereas the algae and eggs followed their original streamlines and thus were collected at the outlet 1 (see also Supplementary Movie 3). In five independent repeated experiments with more than 100 tardigrades, the average recovery rate was measured to be 95.76% ± 2.42% with an average throughput of 400 tardigrades per hour. The average OD values of the samples at the outlet 1 and 2 were 0.1736 (Figure 4(c1)) and 0.0203 (Figure 4(c2)), respectively. Considering the average OD value of the filtered tardigrade sample was 0.512, the algae impurity was 3.96% ± 1.67% in the sample collected at the outlet 2 after the acoustofluidic tardigrade separation.

## 4. DISCUSSION

When the surface acoustic waves (SAWs) interacted with the fluid inside the microchannel, these waves were transformed into compressional waves (CWs) and propagated through the fluid at a Rayleigh angle $\theta_R = \sin^{-1}(c_f/c_s) \approx 22°$, where $c_f$ and $c_s$ are the sound speeds inside the fluid and the substrate, respectively (Collins, Alan, & Neild, 2014; Fakhfouri *et al.*, 2018). When micro-objects are suspended in the microchannel and exposed to the CWs in the path of the wave propagation, these waves are scattered in an inhomogeneous manner at the interface between the liquid and the object; as a result, the suspended microscale objects experience the acoustic radiation force (ARF) in the wave propagation direction. The magnitude of the ARF acting on the micro-objects strongly depends on the dimensionless Helmholtz number, $\kappa = \pi d_m/\lambda_f$, where $d_m$ is the representative length of the objects and $\lambda_f$ is the acoustic wavelength. If $d_m > \lambda_f$, asymmetric wave scattering off the micro-object results in a significant ARF in the direction of the wave propagation whereas the ARF acting on small objects compared to the acoustic wavelength ($d_m < \lambda_f$) is negligible because of homogeneous wave scattering (Skowronek *et al.*, 2013; Skowronek, Rambach, & Franke, 2015; Destgeer *et al.*, 2014a; Devendran *et al.*, 2016). Considering that the average length of the tardigrades that we targeted to manipulate was approximately 100 μm in our experiments, we used interdigital transducers (IDTs) with their resonant frequencies above 36 MHz such that the tardigrades experienced the ARF strong enough to overcome the flow-induced drag force. The 72 MHz SAWs were used in the experiments above to prevent standing SAWs formed in the vertical direction within the parallel-type acoustofluidic device. As shown in Figure 3(b), 72 MHz SAWs, whose wavelength was smaller than the tardigrades but larger than the eggs and algae, imparted the ARF selectively to the tardigrades. As a consequence, only the tardigrade remained standstill while the other components of the tardigrade sample after microfluidic filtration passed freely through the acoustic field. As previously reported (Wiklund, 2012; Li *et al.*, 2015), the ARF-based acoustofluidic manipulation was found to be biocompatible and imparted no harmful stresses to the tardigrades as they moved actively (see also Supplementary Movie 2) and successfully reproduced.

The average recovery rate of the acoustofluidic tardigrade separation was above 95%. A few tardigrades did not experience the ARF enough to be deflected toward the outlet 2 for the following reasons. First, a few tardigrades may have passed through the acoustic anechoic region, where the acoustic field was not effective, due to unintended flow disturbance induced by pressure fluctuations and clogging within the microchannel. Second, the characteristic length of some tardigrades was smaller than the acoustic wavelength ($\lambda_f = 80$ μm for 45 MHz SAWs), so insignificant acoustic radiation occurred around the tardigrades. Third, when multiple tardigrades were passing through the acoustic field close to each other, insufficient magnitude of the acoustic field was applied to the tardigrade located farther from the IDT, leading to the insignificant wave scattering. In the acoustofluidic tardigrade separation, the average time required to collect 100 tardigrades was measured to be 15 min (approximately 400 tardigrades per hour). However, the throughput can be varied depending on the number of tardigrades per unit sample volume injected into the active separation device. In addition, the algae impurity of the tardigrade sample after acoustofluidic separation was measured to be slightly less than 4% based on the optical density (OD) measurement. A very small amount of the algae and eggs in the sample flow were unintendedly separated along with the tardigrades for the following reasons. First, the acoustic streaming flow-induced drag force was imparted to the eggs and algae. Second, unintended flow disturbance caused the tardigrade sample flow to be imperfectly guided to the outlet 1 by the sheath flows. Third, some algae attached to the tardigrades were separated together with the tardigrades that were affected by the SAW-based ARF. The acoustofluidic tardigrade separation device can be easily switched on and off by simply controlling the electrical signal applied to the IDT. The on-demand control of the device can be utilized to prepare the tardigrade samples with the desired amount of the tardigrades.

## 5. CONCLUSION

We developed a two-step tardigrade isolation method to separate tardigrades from raw culture samples in a continuous, automated, contactless, on-demand manner. In the passive microfluidic filtration step, large algal clusters were removed by an array of 218 traps composed of four micropillars, with a significantly decreased OD value by 77% on average, to facilitate the acoustofluidic tardigrade separation. The effects of the acoustic waves on the tardigrade sample were examined in a parallel acoustofluidic device. In our experiments, tardigrades were found to experience the SAW-based ARF that was significant enough to overcome the flow-induced drag force acting on the tardigrades whereas the eggs and algae remained unaffected by the acoustic field. Biocompatibility of the acoustic waves applied to the tardigrades was confirmed by active movement and reproduction of the tardigrades after being exposed to the acoustic field. In the active tardigrade separation device, the passively filtered tardigrade sample was injected to the cross-type acoustofluidic device. We demonstrated that the tardigrades could be selectively deflected by the SAW-induced ARF and thus collected at a separated outlet at a recovery rate of 95.76% and an algae impurity of 3.96%. The proposed acousto-microfluidic approach to tardigrade separation is expected to aid tardigrade research and may be further extended to the selective isolation of the tardigrades with different age groups.

## ACKNOWLEDGEMENTS

This work was supported by the National Research Foundation of Korea (NRF) (Grant No. 2019022966), the

# REFERENCES

Ahmad R, Destgeer G, Afzal M, Park J, Ahmed H, Jung JH, Park K, Yoon TS, Sung HJ. 2017. Acoustic wave-driven functionalized particles for aptamer-based target biomolecule separation. *Analytical Chemistry* 89: 13313–13319.

Ahmed H, Destgeer G, Park J, Afzal M, Sung HJ. 2018. Sheathless focusing and separation of microparticles using tilted-angle traveling surface acoustic waves. *Analytical Chemistry* 90: 8546–8552.

Ahmed H, Destgeer G, Park J, Jung JH, Ahmad R, Park K, Sung HJ. 2017a. A pumpless acoustofluidic platform for size-selective concentration and separation of microparticles. *Analytical Chemistry* 89: 13575–13581.

Ahmed H, Destgeer G, Park J, Jung JH, Sung HJ. 2017b. Vertical hydrodynamic focusing and continuous acoustofluidic separation of particles via upward migration. *Advanced Science* 5: 1700285.

Ahmed H, Park J, Destgeer G, Afzal M, Sung HJ. 2019. Surface acoustic wave-based micromixing enhancement using a single interdigital transducer. *Applied Physics Letters* 114: 043702.

Alnaimat F, Dagher S, Mathew B, Hilal-Alnqbi A, Khashan S. 2018. Microfluidics based magnetophoresis: a review. *The Chemical Record* 18: 1596–1612.

Altiero T, Giovannini I, Guidetti R, Rebecchi L. 2015. Life history traits and reproductive mode of the tardigrade Acutuncus antarcticus under laboratory conditions: strategies to colonize the Antarctic environment. *Hydrobiologia* 761: 277–291.

Altiero T, Guidetti R, Caselli V, Cesari M, Rebecchi L. 2011. Ultraviolet radiation tolerance in hydrated and desiccated eutardigrades. *Journal of Zoological Systematics and Evolutionary Research* 49: 104–110.

Altiero T, Rebecchi L. 2001. Rearing tardigrades: results and problems. *Zoologischer Anzeiger - A Journal of Comparative Zoology* 240: 217–221.

Arakawa K. 2016. No evidence for extensive horizontal gene transfer from the draft genome of a tardigrade. *Proceedings of the National Academy of Sciences* 113: E3057.

Arakawa K, Yoshida Y, Tomita M. 2016. Genome sequencing of a single tardigrade Hypsibius dujardini individual. *Scientific Data* 3: 160063.

Beltrán-Pardo E, Jönsson KI, Harms-Ringdahl M, Haghdoost S, Wojcik A. 2015. Tolerance to gamma radiation in the tardigrade Hypsibius dujardini from embryo to adult correlate inversely with cellular proliferation. *PLoS One* 10: e0133658.

Bemm F, Weiß CL, Schultz J, Förster F. 2016. Genome of a tardigrade: horizontal gene transfer or bacterial contamination? *Proceedings of the National Academy of Sciences* 113: E3054-6.

Van Bezooijen J. 2006. *Methods and techniques for nematology*. Wageningen, The Netherlands: Wageningen University.

Bhagat AAS, Bow H, Hou HW, Tan SJ, Han J, Lim CT. 2010. Microfluidics for cell separation. *Medical & Biological Engineering & Computing* 48: 999–1014.

Boothby TC, Tenlen JR, Smith FW, Wang JR, Patanella KA, Osborne Nishimura E, Tintori SC, Li Q, Jones CD, Yandell M, Messina DN, Glasscock J, Goldstein B. 2015. Evidence for extensive horizontal gene transfer from the draft genome of a tardigrade. *Proceedings of the National Academy of Sciences* 112: 15976–15981.

Choi S, Song S, Choi C, Park JK. 2008. Sheathless focusing of microbeads and blood cells based on hydrophoresis. *Small* 4: 634–641.

Collins DJ, Alan T, Neild A. 2014. The particle valve: on-demand particle trapping, filtering, and release from a microfabricated polydimethylsiloxane membrane using surface acoustic waves. *Applied Physics Letters* 105: 033509.

Collins DJ, O'Rorke R, Devendran C, Ma Z, Han J, Neild A, Ai Y. 2018. Self-aligned acoustofluidic particle focusing and patterning in microfluidic channels from channel-based acoustic waveguides. *Physical Review Letters* 120: 074502.

Degma P, Katina S, Sabatovičová L. 2011. Horizontal distribution of moisture and Tardigrada in a single moss cushion. *Journal of Zoological Systematics and Evolutionary Research* 49: 71–77.

Destgeer G, Cho H, Ha BH, Jung JH, Park J, Sung HJ. 2016. Acoustofluidic particle manipulation inside a sessile droplet: four distinct regimes of particle concentration. *Lab on a Chip* 16: 660–667.

Destgeer G, Ha BH, Jung JH, Sung HJ. 2014a. Submicron separation of microspheres via travelling surface acoustic waves. *Lab on a Chip* 14: 4665–4672.

Destgeer G, Ha BH, Park J, Jung JH, Alazzam A, Sung HJ. 2015. Microchannel anechoic corner for size-selective separation and medium exchange via traveling surface acoustic waves. *Analytical Chemistry* 87: 4627–4632.

Destgeer G, Hashmi A, Park J, Ahmed H, Afzal M, Sung HJ. 2019. Microparticle self-assembly induced by travelling surface acoustic waves. *RSC Advances* 9: 7916–7921.

Destgeer G, Im S, Ha BH, Jung JH, Ansari MA, Sung HJ. 2014b. Adjustable, rapidly switching microfluidic gradient generation using focused travelling surface acoustic waves. *Applied Physics Letters* 104: 023506.

Destgeer G, Lee KH, Jung JH, Alazzam A, Sung HJ. 2013. Continuous separation of particles in a PDMS microfluidic channel via travelling surface acoustic waves (TSAW). *Lab on a Chip* 13: 4210–6.

Destgeer G, Sung HJ. 2015. Recent advances in microfluidic actuation and micro-object manipulation via surface acoustic waves. *Lab on a Chip* 15: 2722–2738.

Devendran C, Albrecht T, Brenker J, Alan T, Neild A. 2016. The importance of travelling wave components in standing surface acoustic wave (SSAW) systems. *Lab on a Chip* 16: 3756–3766.

Ding X, Li P, Lin SCS, Stratton ZS, Nama N, Guo F, Slotcavage D, Mao X, Shi J, Costanzo F, Huang TJ. 2013. Surface acoustic wave microfluidics. *Lab on a Chip* 13: 3626–49.

Ding X, Lin SCS, Kiraly B, Yue H, Li S, Chiang IK, Shi J, Benkovic SJ, Huang TJ. 2012. On-chip manipulation of single microparticles, cells, and organisms using surface acoustic waves. *Proceedings of the National Academy of Sciences* 109: 11105–9.

Fakhfouri A, Devendran C, Albrecht T, Collins DJ, Winkler A, Schmidt H, Neild A. 2018. Surface acoustic wave diffraction driven mechanisms in microfluidic systems. *Lab on a Chip* 18: 2214–2224.

Friend J, Yeo L. 2010. Fabrication of microfluidic devices using polydimethylsiloxane. *Biomicrofluidics* 4: 026502.

Gabriel WN, McNuff R, Patel SK, Gregory TR, Jeck WR, Jones CD, Goldstein B. 2007. The tardigrade Hypsibius dujardini, a new model for studying the evolution of development. *Developmental Biology* 312: 545–59.

Gasiorek P, Stec D, Morek W, Michalczyk Ł. 2018. An integrative redescription of Hypsibius dujardini (Doyère, 1840), the nominal taxon for Hypsibioidea (Tardigrada: Eutardigrada). *Zootaxa* 4415(1): 45–75.

Giovannini I, Altiero T, Guidetti R, Rebecchi L. 2018. Will the Antarctic tardigrade Acutuncus antarcticus be able to withstand environmental stresses related to global climate change? *Journal of Experimental Biology* 221: 160622.

Goldstein B, King N. 2016. The future of cell biology: emerging model organisms. *Trends in Cell Biology* 26: 818–824.

Gross V, Minich I, Mayer G. 2017. External morphogenesis of the tardigrade Hypsibius dujardini as revealed by scanning electron microscopy. *Journal of Morphology* 278: 563–573.

Guidetti R, Altiero T, Rebecchi L. 2011. On dormancy strategies in tardigrades. *Journal of Insect Physiology* 57: 567–576.

Ha BH, Park J, Destgeer G, Jung JH, Sung HJ. 2015. Generation of dynamic free-form temperature gradients in a disposable microchip. *Analytical Chemistry* 87: 11568–11574.

Hashimoto T, Horikawa DD, Saito Y, Kuwahara H, Kozuka-Hata H, Shin-I T, Minakuchi Y, Ohishi K, Motoyama A, Aizu T, Enomoto A, Kondo K, Tanaka S, Hara Y, Koshikawa S, Sagara H, Miura T, Yokobori S ichi, Miyagawa K, Suzuki Y, Kubo T, Oyama M, Kohara Y, Fujiyama A, Arakawa K, Katayama T, Toyoda A, Kunieda T. 2016. Extremotolerant tardigrade genome and improved radiotolerance of human cultured cells by tardigrade-unique protein. *Nature Communications* 7: 12808.

Hengherr S, Worland MR, Reuner A, Brümmer F, Schill RO. 2009a. High☐temperature tolerance in anhydrobiotic tardigrades is limited by glass transition. *Physiological and Biochemical Zoology* 82: 749–755.

Hengherr S, Worland MR, Reuner A, Brümmer F, Schill RO. 2009b. Freeze tolerance, supercooling points and ice formation: comparative studies on the subzero temperature survival of limno-terrestrial tardigrades. *Journal of Experimental Biology* 212: 802–7.

Horikawa DD, Cumbers J, Sakakibara I, Rogoff D, Leuko S, Harnoto R, Arakawa K, Katayama T, Kunieda T, Toyoda A, Fujiyama A, Rothschild LJ. 2013. Analysis of DNA repair and protection in the tardigrade Ramazzottius varieornatus and Hypsibius dujardini after exposure to UVC radiation. *PLoS One* 8: e64793.

Horikawa DD, Kunieda T, Abe W, Watanabe M, Nakahara Y, Yukuhiro F, Sakashita T, Hamada N, Wada S, Funayama T, Katagiri C, Kobayashi Y, Higashi S, Okuda T. 2008. Establishment of a rearing system of the extremotolerant tardigrade Ramazzottius varieornatus: a new model animal for astrobiology. *Astrobiology* 8: 549–556.

Hur SC, Tse HTK, Di Carlo D. 2010. Sheathless inertial cell ordering for extreme throughput flow cytometry. *Lab on a Chip* 10: 274–280.

Jönsson KI. 2007. Tardigrades as a potential model organism in space research. *Astrobiology* 7: 757–766.

Jönsson KI, Rabbow E, Schill RO, Harms-Ringdahl M, Rettberg P. 2008. Tardigrades survive exposure to space in low earth orbit. *Current Biology* 18: R729–R731.

Jørgensen A, Kristensen RM, Møbjerg N. 2018. Phylogeny and integrative taxonomy of Tardigrada. In: Schill RO, ed. *Water Bears: The Biology of Tardigrades. Zoological Monographs vol. 2*. Cham, Switzerland: Springer, .

Jung JH, Lee KH, Lee KS, Ha BH, Oh YS, Sung HJ. 2014. Optical separation of droplets on a microfluidic platform. *Microfluidics and Nanofluidics* 16: 635–644.

Kosztyła P, Stec D, Morek W, Gąsiorek P, Zawierucha K, Michno K, Ufir K, Małek D, Hlebowicz K, Laska A, Dudziak M, Frohme M, Prokop ZM, Kaczmarek Ł, Michalczyk Ł. 2016. Experimental taxonomy confirms the environmental stability of morphometric traits in a taxonomically challenging group of microinvertebrates. *Zoological Journal of the Linnean Society* 178(4): 765–775.

Koutsovoulos G, Kumar S, Laetsch DR, Stevens L, Daub J, Conlon C, Maroon H, Thomas F, Aboobaker AA, Blaxter M. 2016. No evidence for extensive horizontal gene transfer in the genome of the tardigrade *Hypsibius dujardini*. *Proceedings of the National Academy of Sciences* 113: 5053–5058.

Li P, Mao Z, Peng Z, Zhou L, Chen Y, Huang PH, Truica CI, Drabick JJ, El-Deiry WS, Dao M, Suresh S, Huang TJ. 2015. Acoustic separation of circulating tumor cells. *Proceedings of the National Academy of Sciences* 112: 4970–4975.

Ma Z, Zhou Y, Collins DJ, Ai Y. 2017. Fluorescence activated cell sorting via a focused traveling surface acoustic beam. *Lab on a Chip* 17: 3176–3185.

Martin C, Gross V, Hering L, Tepper B, Jahn H, de Sena Oliveira I, Stevenson PA, Mayer G. 2017. The nervous and visual systems of onychophorans and tardigrades: learning about arthropod evolution from their closest relatives. *Journal of Comparative Physiology A* 203: 565–590.

Mathew B, Alazzam A, Destgeer G, Sung HJ. 2016. Dielectrophoresis based cell switching in continuous flow microfluidic devices. *Journal of Electrostatics* 84: 63–72.

Møbjerg N, Halberg KA, Jørgensen A, Persson D, Bjørn M, Ramløv H, Kristensen RM. 2011. Survival in extreme environments - on the current knowledge of adaptations in tardigrades. *Acta Physiologica* 202: 409–420.

Nam J, Jang WS, Lim CS. 2018. Micromixing using a conductive liquid-based focused surface acoustic wave (CL-FSAW). *Sensors and Actuators B: Chemical* 258: 991–997.

Nelson DR. 2002. Current status of the Tardigrada: evolution and ecology. *Integrative and Comparative Biology* 42: 652–659.

Ono F, Mori Y, Takarabe K, Fujii A, Saigusa M, Matsushima Y, Yamazaki D, Ito E, Galas S, Saini NL. 2016. Effect of ultra-high pressure on small animals, tardigrades and Artemia. *Cogent Physics* 3: 1167575.

Ono F, Saigusa M, Uozumi T, Matsushima Y, Ikeda H, Saini NL, Yamashita M. 2008. Effect of high hydrostatic pressure on to life of the tiny animal tardigrade. *Journal of Physics and Chemistry of Solids* 69: 2297–2300.

Park J, Destgeer G, Kim H, Cho Y, Sung HJ. 2018a. In-droplet microparticle washing and enrichment using surface acoustic wave-driven acoustic radiation force. *Lab on a Chip* 18: 2936–2945.

Park J, Jung JH, Destgeer G, Ahmed H, Park K, Sung HJ. 2017a. Acoustothermal tweezer for droplet sorting in a disposable microfluidic chip. *Lab on a Chip* 17: 1031–1040.

Park J, Jung JH, Park K, Destgeer G, Ahmed H, Ahmad R, Sung HJ. 2018b. On-demand acoustic droplet splitting and steering in a disposable microfluidic chip. *Lab on a Chip* 18: 422–432.

Park K, Park J, Jung JH, Destgeer G, Ahmed H, Sung HJ. 2017b. In-droplet microparticle separation using travelling surface acoustic wave. *Biomicrofluidics* 11: 064112.

Pilato G, Binda MG. 2001. Biogeography and limno-terrestrial tardigrades: are they truly incompatible binomials? *Zoologischer Anzeiger - A Journal of Comparative Zoology* 240: 511–516.

Rebecchi L, Altiero T, Guidetti R, Cesari M, Bertolani R, Negroni M, Rizzo AM. 2009. Tardigrade resistance to space effects: first results of experiments on the LIFE-TARSE mission on FOTON-M3 (September 2007). *Astrobiology* 9: 581–591.

Sajeesh P, Sen AK. 2014. Particle separation and sorting in microfluidic devices: a review. *Microfluidics and Nanofluidics* 17: 1–52.

Sands CJ, Convey P, Linse K, McInnes SJ. 2008. Assessing meiofaunal variation among individuals utilising morphological and molecular approaches: an example using the Tardigrada. *BMC Ecology* 8: 7.

Schokraie E, Warnken U, Hotz-Wagenblatt A, Grohme MA, Hengherr S, Forster F, Schill RO, Frohme M, Dandekar T, Schnolzer M. 2012. Comparative proteome analysis of Milnesium tardigradum in early embryonic state versus adults in active and anhydrobiotic state. *PLoS One* 7: e45682.

Seki K, Toyoshima M. 1998. Preserving tardigrades under pressure. *Nature* 395: 853–854.

Shi J, Huang H, Stratton Z, Huang Y, Huang TJ. 2009. Continuous particle separation in a microfluidic channel via standing surface acoustic waves (SSAW). *Lab on a Chip* 9: 3354–3359.

Skowronek V, Rambach RW, Franke T. 2015. Surface acoustic wave controlled integrated band-pass filter. *Microfluidics and Nanofluidics* 19: 335–341.

Skowronek V, Rambach RW, Schmid L, Haase K, Franke T. 2013. Particle deflection in a poly(dimethylsiloxane) microchannel using a propagating surface acoustic wave: size and frequency dependence. *Analytical Chemistry* 85: 9955–9.

Smith FW, Boothby TC, Giovannini I, Rebecchi L, Jockusch EL, Goldstein B. 2016. The compact body plan of tardigrades evolved by the loss of a large body region. *Current Biology* 26: 224–229.

Suzuki AC. 2003. Life history of Milnesium tardigradum Doyère (tardigrada) under a rearing environment. *Zoological Science* 20: 49–57.

Tanaka S, Tanaka J, Miwa Y, Horikawa DD, Katayama T, Arakawa K, Toyoda A, Kubo T, Kunieda T. 2015. Novel mitochondria-targeted heat-soluble proteins identified in the anhydrobiotic tardigrade improve osmotic tolerance of human cells (JR Battista, Ed.).

*PLoS One* 10: e0118272.

Tenlen JR, McCaskill S, Goldstein B. 2013. RNA interference can be used to disrupt gene function in tardigrades. *Development Genes and Evolution* 223: 171–181.

Tsujimoto M, Imura S, Kanda H. 2016. Recovery and reproduction of an Antarctic tardigrade retrieved from a moss sample frozen for over 30 years. *Cryobiology* 72: 78–81.

Tsujimoto M, Suzuki AC, Imura S. 2015. Life history of the Antarctic tardigrade, Acutuncus antarcticus, under a constant laboratory environment. *Polar Biology* 38: 1575–1581.

Weibel DB, Whitesides GM, Basu A, Schneider J. 2006. Applications of microfluidics in chemical biology. *Current Opinion in Chemical Biology* 10: 584–591.

Weronika E, Łukasz K. 2017. Tardigrades in space research - past and future. *Origins of Life and Evolution of Biospheres* 47: 545–553.

Wiklund M. 2012. Acoustofluidics 12: biocompatibility and cell viability in microfluidic acoustic resonators. *Lab on a Chip* 12: 2018–2028.

Winkler A, Harazim S, Collins DJ, Brünig R, Schmidt H, Menzel SB. 2017. Compact SAW aerosol generator. *Biomedical Microdevices* 19: 9.

Wyatt Shields IV C, Reyes CD, López GP. 2015. Microfluidic cell sorting: a review of the advances in the separation of cells from debulking to rare cell isolation. *Lab on a Chip* 15: 1230–1249.

Yamaguchi A, Tanaka S, Yamaguchi S, Kuwahara H, Takamura C, Imajoh-Ohmi S, Horikawa DD, Toyoda A, Katayama T, Arakawa K, Fujiyama A, Kubo T, Kunieda T. 2012. Two novel heat-soluble protein families abundantly expressed in an anhydrobiotic tardigrade. *PLoS One* 7: e44209.

Yoshida Y, Koutsovoulos G, Laetsch DR, Stevens L, Kumar S, Horikawa DD, Ishino K, Komine S, Kunieda T, Tomita M, Blaxter M, Arakawa K. 2017. Comparative genomics of the tardigrades Hypsibius dujardini and Ramazzottius varieornatus. *PLoS Biology* 15: e2002266.